\documentclass{article}
\usepackage{amsmath}
\usepackage{amssymb}
\usepackage{amsfonts}
\usepackage{tikz}
\usepackage{latexsym,longtable}
\usepackage[colorlinks,linkcolor=blue,anchorcolor=blue,citecolor=green,CJKbookmarks=True]{hyperref}
\usepackage{tikz}

\newcommand{\C}{{\cal C}}

\newcommand{\G}{{\cal G}}

\newtheorem{theorem}{Theorem}[section]

\newtheorem{lemma}[theorem]{Lemma}

\def\whitebox{{\hbox{\hskip 1pt
 \vrule height 6pt depth 1.5pt
 \lower 1.5pt\vbox to 7.5pt{\hrule width
    3.2pt\vfill\hrule width 3.2pt}%
 \vrule height 6pt depth 1.5pt
 \hskip 1pt } }}
\def\qed{\ifhmode\allowbreak\else\nobreak\fi\hfill\quad\nobreak
     \whitebox\medbreak}
\newcommand{\proof}{\noindent{\it Proof:}\ }

\newcommand{\ignore}[1]{}
\begin{document}

\title{\bf Quantum Latin squares of order $6m$ with all possible cardinalities \thanks{Research supported by  NSFC Grant 12271390.}}

\author{
	{\small Ying Zhang,  Lijun Ji}\\
	{\small Department of Mathematics}, {\small Soochow University},
	{\small Suzhou  215006, China}\\
	\texttt{yingzhangzy2025@163.com; jilijun@suda.edu.cn}\\
}

\date{}
\maketitle
\begin{abstract}
	\noindent\noindent
	A quantum Latin square of order $n$ (denoted as QLS$(n)$) is an $n\times n$ array whose entries are unit column vectors from the $n$-dimensional Hilbert space $\mathcal{H}_n$, such that each row and column forms an orthonormal basis. Two unit vectors $|u\rangle, |v\rangle\in  \mathcal{H}_n$ are regarded as identical if there exists a real number $\theta$ such that
	$|u\rangle=e^{i\theta}|v\rangle$; otherwise, they are considered distinct.
	The cardinality $c$ of a QLS$(n)$ is the number of distinct vectors in the
	array. In this note,we use sub-QLS$(6)$ to prove that for any integer $m\geq 2$ and any $c\in [6m,36m^2]\setminus \{6m+1\}$, there is a QLS$(6m)$ with cardinality $c$.


\end{abstract}

\section{Introduction}

A quantum Latin square of order $n$, denoted as QLS$(n)$, is an $n\times n$ array whose entries are unit column vectors from the $n$-dimensional Hilbert space $\mathcal{H}_n$, such that each row and column forms an orthonormal basis. For any classical Latin square of order $n$ over $\{0,\ldots,n-1\}$, it is possible to obtain a quantum Latin square of order $n$ by interpreting a number $i\in \{0,1,\ldots,n-1\}$ as a column vector $|i\rangle$ in $\mathcal{H}_n$. Specifically,
$|i\rangle$ is a unit vector with its $(i+1)$th component equal to 1, and $\{|0\rangle, |1\rangle, \ldots, |n-1\rangle\}$ is an
orthonormal basis for $\mathcal{H}_n$, referred to as computational basis.
In 2016, Musto and Vicary introduced
quantum Latin squares as a quantum-theoretic generalization of classical Latin squares,
showing their utility in constructing unitary error bases (UEBs) \cite{MV2016}.

In quantum theory \cite{NC2010}, for any two unit vectors $|u\rangle, |v\rangle\in  \mathcal{H}_n$, if there exists a real number $\theta$ such that
$$|u\rangle=e^{i\theta}|v\rangle,$$
then $|u\rangle$ and $|v\rangle$ are considered to represent the same quantum state and are thus regarded
as identical; otherwise, they are considered distinct. 
The cardinality $c$ of a QLS$(n)$ is the number of distinct vectors in the
array. Since a LS$(n)$ always exists, there is a QLS$(n)$ with cardinality $n$. Clearly the cardinality $c$ of a QLS$(n)$ satisfies that $n\leq c\leq  n^2$. A QLS$(n)$ is called apparently quantum if $c = n$, or genuinely quantum if $n<c\leq n^2$. Genuinely quantum Latin squares of order 2 or 3 cannot exist and the possible cardinalities of a quantum Latin squares of order 4 are 4, 6, 8, and 16 \cite{PWRBZ2021}.

In 2021, Nechita and Pillet \cite{NP2021} proposed the concept of quantum Sudoku, a special
case of quantum Latin squares. An $n^2\times n^2$ quantum Sudoku is a QLS$(n^2)$ partitioned
into $n$ disjoint $n\times n$ blocks, with the additional requirement that each block forms an
orthonormal basis of $\mathcal{H}_{n^2}$. In the same year, Paczos et al. \cite{PWRBZ2021} introduced the cardinality
measure for quantum Latin squares, proving the existence of $n^2\times n^2$ quantum Sudoku with maximal cardinality for any positive integer $n$.  Recently, 
Zhang and Cao has almost determined the existence of QLS$(n)$ with maximal cardinality $n^2$ with a few possible exceptions and two definite exceptions $n=2,3$ \cite{ZCPreprint}. For further reading on quantum
theory and quantum Latin squares, the reader is referred to \cite{CLV2024,GRDZ2018,LNV2023,LW2018,RRKL2023,ZTFZ2023,ZTZF2022}.
In this paper, we construct QLS$(4m)$s with all possible cardinalities. 

The cardinality $c$ of a QLS$(n)$ is the number of distinct vectors in the
array. Since a LS$(n)$ always exists, there is a QLS$(n)$ with cardinality $n$. Clearly the cardinality $c$ of a QLS$(n)$ satisfies that $n\leq c\leq  n^2$. Zhang and Cao has almostly determined the existence of QLS$(n)$ with maximal cardinality $n^2$ with a few possible exceptions and two difinite exceptions $n=2,3$ \cite{ZCPreprint}. Recently, these possible exceptions were determined by Zang et al \cite{ZZTS}. 

\begin{lemma}\label{n+1}
For any integer $n\geq 2$, there does not exist a QLS$(n)$ with cardinality $n+1$.	
\end{lemma}

\begin{theorem}[\cite{ZZTS}]\label{MQLS(n)}
	For any integer $n\geq 4$, there is a QLS$(n)$ with maximal cardinality $n^2$.
\end{theorem}

\begin{theorem}[\cite{ZWJPreprint}]\label{QLS(4m)}
	For any integer $m\geq 2$ and any integer $c\in [4m,16m^2]\setminus \{4m+1\}$, there is a QLS$(4m)$ over $\mathcal{H}_m\otimes\mathcal{H}_2\otimes \mathcal{H}_2$ with cardinality $c$.
\end{theorem}

In this note, we construct QLS$(6m)$ with all possible cardinalities. 

\section{Main result}

An $m\times n$ row-quantum Latin rectangle is an array with $m$ rows and $n$ columns whose entries are unit vectors from $n$-dimensional Hilbert space $\mathcal{H}_n$, and such that
the elements in each row are mutually orthogonal.

Let $U$ and $V$ denote an $m\times n$ row-quantum Latin rectangle with cardinality $c_1$ in $\mathcal{H}_n$ and
an $n\times m$ row-quantum Latin rectangle with cardinality $c_2$ in $\mathcal{H}_m$, respectively. Let $|u_{i,j}\rangle$ and $|v_{k,l}\rangle$
represent the elements in the $i$th row and $j$th column of $U$, and the $k$th row and $l$th column
of $V$, respectively. Let $W$ be a matrix of order $mn$, which is divided into $mn$ blocks of size $n\times m$. Denote by
$|w_{i,j;k,l}\rangle$ the element located in the $k$th row and $l$th column of the $(i,j)$th block in the
matrix $W$ and define
\begin{equation}\label{matrixtensorproduct}
|w_{i,j;k,l}\rangle =|u_{i,j+k}\rangle \otimes |v_{j,i+l}\rangle
\end{equation}
where the addition $j + k$ is modulo $n$ and $i + l$ is modulo $m$. 

\begin{lemma}{\rm \cite[Theorem 3.4]{ZCPreprint}}\label{ProductConstruction}
Let $U$ and $V$ denote an $m\times n$ row-quantum Latin rectangle with cardinality $c_1$ in $\mathcal{H}_n$ and
an $n\times m$ row-quantum Latin rectangle with cardinality $c_2$ in $\mathcal{H}_m$, respectively. Then the $mn\times mn$ matrix $W$ defined by Equation $(\ref{matrixtensorproduct})$ is a QLS$(mn)$ with cardinality $c_1c_2$.
\end{lemma}

For simplicity, $|i\rangle\otimes |j\rangle$ is written as $|ij\rangle$ for $i\in \{0,1\}$, $j\in \{0,1,2\}$. For any  $a\in \mathbb{R}$, define 
\begin{equation}\label{A_aB_a}
\begin{array}{l}
A_a=\frac{1}{\sqrt{1+a^2}}\begin{pmatrix} |00\rangle+a|01\rangle & -a|00\rangle+|01\rangle\\
	-a|00\rangle+|01\rangle & |00\rangle+a|01\rangle\\
\end{pmatrix},\smallskip \\ B_a=\frac{1}{\sqrt{1+a^2}}\begin{pmatrix} |10\rangle+a|11\rangle & -a|10\rangle+|11\rangle\\
	-a|10\rangle+|11\rangle & |10\rangle+a|11\rangle\\
\end{pmatrix},\smallskip \\
C_a=\frac{1}{\sqrt{1+a^2}}\begin{pmatrix} |02\rangle+a|12\rangle & -a|02\rangle+|12\rangle\\
		-a|02\rangle+|12\rangle & |02\rangle+a|12\rangle\\
\end{pmatrix},\smallskip \\
D_a=\frac{1}{\sqrt{1+a^2}}\begin{pmatrix} \sqrt{1+a^2}|00\rangle&|01\rangle+a|02\rangle & -a|01\rangle+|02\rangle\\
-a|01\rangle+|02\rangle &\sqrt{1+a^2}|00\rangle& |01\rangle+a|02\rangle\\
|01\rangle+a|02\rangle&-a|01\rangle+|02\rangle&\sqrt{1+a^2}|00\rangle\\\end{pmatrix},\smallskip \\
E_a=\frac{1}{\sqrt{1+a^2}}\begin{pmatrix} \sqrt{1+a^2}|10\rangle&|11\rangle+a|12\rangle & -a|11\rangle+|12\rangle\\
-a|11\rangle+|12\rangle &\sqrt{1+a^2}|10\rangle& |11\rangle+a|12\rangle\\
|11\rangle+a|12\rangle&-a|11\rangle+|12\rangle&\sqrt{1+a^2}|10\rangle\\
\end{pmatrix},\smallskip \\

F_1=\frac{1}{\sqrt{3}}\begin{pmatrix} |00\rangle+\omega|01\rangle+\omega^2|02\rangle&|00\rangle+\omega^2|01\rangle+\omega|02\rangle & |00\rangle+|01\rangle+|02\rangle\\
|00\rangle+\omega^2|01\rangle+\omega|02\rangle & |00\rangle+|01\rangle+|02\rangle&|00\rangle+\omega|01\rangle+\omega^2|02\rangle\\
|00\rangle+|01\rangle+|02\rangle&|00\rangle+\omega|01\rangle+\omega^2|02\rangle&|00\rangle+\omega^2|01\rangle+\omega|02\rangle\\\end{pmatrix},\smallskip \\
F_2 = \begin{pmatrix} 
	-\frac{1}{2}|00\rangle - \frac{\sqrt{3}}{2}|01\rangle & \frac{\sqrt{3}}{6}|00\rangle - \frac{1}{6}|01\rangle - \frac{\sqrt{6}}{3}|02\rangle & \frac{\sqrt{3}}{6}|00\rangle - \frac{1}{6}|01\rangle + \frac{\sqrt{6}}{3}|02\rangle \\
	\frac{\sqrt{3}}{6}|00\rangle - \frac{1}{6}|01\rangle - \frac{\sqrt{6}}{3}|02\rangle & \frac{\sqrt{3}}{6}|00\rangle - \frac{1}{6}|01\rangle + \frac{\sqrt{6}}{3}|02\rangle & -\frac{1}{2}|00\rangle - \frac{\sqrt{3}}{2}|01\rangle \\
	\frac{\sqrt{3}}{6}|00\rangle - \frac{1}{6}|01\rangle + \frac{\sqrt{6}}{3}|02\rangle & -\frac{1}{2}|00\rangle - \frac{\sqrt{3}}{2}|01\rangle & \frac{\sqrt{3}}{6}|00\rangle - \frac{1}{6}|01\rangle - \frac{\sqrt{6}}{3}|02\rangle 
\end{pmatrix},
\smallskip \\
G_1=\frac{1}{\sqrt{3}}\begin{pmatrix} |10\rangle+\omega|11\rangle+\omega^2|12\rangle&|10\rangle+\omega^2|11\rangle+\omega|12\rangle & |10\rangle+|11\rangle+|12\rangle\\
	|10\rangle+\omega^2|11\rangle+\omega|12\rangle & |10\rangle+|11\rangle+|12\rangle&|10\rangle+\omega|11\rangle+\omega^2|12\rangle\\
	|00\rangle+|11\rangle+|12\rangle&|10\rangle+\omega|11\rangle+\omega^2|12\rangle&|10\rangle+\omega^2|11\rangle+\omega|12\rangle\\
\end{pmatrix},
\end{array}	
\end{equation}
where $\omega$ is a primitive cubic root of unity.

Clearly, each $A_a$ is a QLS$(2)$ over  $\mathcal{L}(|00\rangle,|01\rangle)=\{c_1|00\rangle+c_2|01\rangle\colon c_1,c_2\in \mathbb{C}\}$,  $B_a$ is a QLS$(2)$ over $\mathcal{L}(|10\rangle,|11\rangle)=\{c_1|10\rangle+c_2|11\rangle\colon c_1,c_2\in \mathbb{C}\}$,  $C_a$ is a QLS$(2)$ over $\mathcal{L}(|02\rangle,|12\rangle)=\{c_1|02\rangle+c_2|12\rangle\colon c_1,c_2\in \mathbb{C}\}$, $D_a$ is a QLS$(3)$ over  $\mathcal{L}(|00\rangle,|01\rangle,|02\rangle)=\{c_1|00\rangle+c_2|01\rangle+c_3|02\rangle\colon c_1,c_2,c_3\in \mathbb{C}\}$,  $E_a$ is a QLS$(3)$ over  $\mathcal{L}(|10\rangle,|11\rangle,|12\rangle)=\{c_1|10\rangle+c_2|11\rangle+c_3|12\rangle\colon c_1,c_2,c_3\in \mathbb{C}\}$. Hence, the $6\times 6$ matrix of the form $\begin{pmatrix}
	A_b & B_x & C_r\\
	C_s & A_c & B_y\\
	B_z & C_t & A_d\\
\end{pmatrix}$ with arbitrary $b,c,d,x,y,z,r,s,t\in \mathbb{R}$ is also a QLS$(6)$ in $\mathcal{H}_2\otimes \mathcal{H}_3$ and the $6\times 6$ matrix of the form $\begin{pmatrix}
	D_b & E_x \\
	E_y & D_c\\
\end{pmatrix}$ with arbitrary $b,c,x,y\in \mathbb{R}$ is also a QLS$(6)$ in $\mathcal{H}_2\otimes \mathcal{H}_3$.

\begin{lemma}\label{QLS(6)}
	For $c\in \{6,8,9,10,11,12,14,16,18,20,22,24,26,28,30,36\} $, there is a QLS$(6)$ with cardinality $c$.
\label{lemma:QLS(6)}
\end{lemma}

\proof Let $\begin{pmatrix}
	X_{0,0} & X_{0,1} \\
	X_{1,0} & X_{1,1}\\
\end{pmatrix}$ be a QLS$(4)$  in $\mathcal{L}(|00\rangle,|01\rangle,|10\rangle,|11\rangle)$, where each $X_{i,j}$ is a $2\times 2$ submatrix for $i,j\in \{0,1\}$. We construct a QLS$(6)$ in $\mathcal{H}_2\otimes \mathcal{H}_3$ as follows.  
	$$L_{a,b,d}=\begin{pmatrix} C_a &	X_{0,0}&X_{0,1}\\
		X_{1,0}&C_b&X_{1,1}\\
		X_{0,0}&X_{1,0}&C_d
	\end{pmatrix},$$
	where $a,b,c\in \mathbb{C}$ and $C_a$ is defined in (\ref{A_aB_a}). 
	Because  there is a  QLS$(4)$ with cardinality $c\in\{4,6,8,16\}$, the QLS$(6)$ $L_{a,b,c}$ can have 
	cardinality $c+2x$ where $x=|\{a,b,d\}|$, thereby, $c+2x$ can be an arbitrary even number in $[6,22]\setminus \{16\}$.
	
	Let $U$, $V_1$ and $V_2$ denote three $2\times 2$ row-quantum Latin rectangles with cardinality $4$ in $\mathcal{H}_2$, respectively. 
	\begin{center}
		$U=\begin{pmatrix}
			\vert 0 \rangle & \vert 1 \rangle \\
			\frac{3}{5}\vert 0 \rangle + \frac{4}{5}\vert 1 \rangle & -\frac{4}{5}\vert 0 \rangle + \frac{3}{5}\vert 1 \rangle
		\end{pmatrix}$\\
	   \smallskip	 
		$V_1=\begin{pmatrix}
			\vert 0 \rangle & \vert 1 \rangle \\
			\frac{5}{13}\vert 0 \rangle + \frac{12}{13}\vert 1 \rangle & -\frac{12}{13}\vert 0 \rangle + \frac{5}{13}\vert 1 \rangle
		\end{pmatrix}$\\ \smallskip
		$V_2=\begin{pmatrix}
			\frac{3}{5}\vert 0 \rangle + \frac{4}{5}\vert 1 \rangle & -\frac{4}{5}\vert 0 \rangle + \frac{3}{5}\vert 1 \rangle \smallskip \\
			\frac{5}{13}\vert 0 \rangle + \frac{12}{13}\vert 1 \rangle & -\frac{12}{13}\vert 0 \rangle + \frac{5}{13}\vert 1 \rangle
		\end{pmatrix}$
	\end{center}
	Let $|u_{i,j}\rangle$ and $|v_{k,l}\rangle$
	represent the elements in the $i$th row and $j$th column of $U$, and the $k$th row and $l$th column
	of $V_1$ or $V_2$, respectively.  Let $4\times 4$ matrix $X'$ and $X''$  be defined by Equation $(\ref{matrixtensorproduct})$ with $U$ and $V_1$ or $V_2$, respectively. Then $
	X'=\begin{pmatrix}
		X'_{0,0}&X'_{0,1}\\X'_{1,0}&X'_{1,1}
	\end{pmatrix}$ and $X''=
	\begin{pmatrix}
		X''_{0,0}&X'_{0,1}\\X''_{1,0}&X'_{1,1}
	\end{pmatrix}$ are QLS$(4)$'s with cardinality $16$,
	where $X'_{0,0},X'_{0,1},X'_{1,0},X'_{1,1},X''_{0,0},X''_{1,0}$ are as follows.
	\begin{center}
		$X'_{0,0}=
		\begin{pmatrix}
			\vert 11 \rangle & \vert 10\rangle \\
			-\frac{4}{5}\vert 00 \rangle + \frac{3}{5}\vert 10 \rangle & -\frac{4}{5}\vert 01 \rangle + \frac{3}{5}\vert 11 \rangle
		\end{pmatrix}$
		\\
		$X'_{0,1}=
		\begin{pmatrix}
			-\frac{12}{13}\vert 00 \rangle + \frac{5}{13}\vert 01 \rangle & \frac{5}{13}\vert 00 \rangle + \frac{12}{13}\vert 01 \rangle \smallskip \\
			\frac{3}{13}\vert 00 \rangle + \frac{36}{65}\vert 01 \rangle + \frac{4}{13}\vert 10 \rangle + \frac{48}{65}\vert 11 \rangle & -\frac{36}{65}\vert 00 \rangle + \frac{3}{13}\vert 01 \rangle - \frac{48}{65}\vert 10 \rangle + \frac{4}{13}\vert 11 \rangle
		\end{pmatrix}$
		\\
		$X'_{1,0}=
		\begin{pmatrix}
			\vert 01 \rangle & \vert 00 \rangle \smallskip \\
			\frac{3}{5}\vert 00 \rangle + \frac{4}{5}\vert 10 \rangle & \frac{3}{5}\vert 01 \rangle + \frac{4}{5}\vert 11 \rangle
		\end{pmatrix}$\smallskip \\
		$X'_{1,1}=
		\begin{pmatrix}
			-\frac{12}{13}\vert 10 \rangle + \frac{5}{13}\vert 11 \rangle & \frac{5}{13}\vert 10 \rangle + \frac{12}{13}\vert 11 \rangle \smallskip \\
			-\frac{4}{13}\vert 00 \rangle - \frac{48}{65}\vert 01 \rangle + \frac{3}{13}\vert 10 \rangle + \frac{36}{65}\vert 11 \rangle & \frac{48}{65}\vert 00 \rangle - \frac{4}{13}\vert 01 \rangle - \frac{36}{65}\vert 10 \rangle + \frac{3}{13}\vert 11 \rangle
		\end{pmatrix}$
		
	\end{center}
	\begin{center}
		$X''_{0,0}=
		\begin{pmatrix}
			-\frac{4}{5}\vert 10 \rangle + \frac{3}{5}\vert 11 \rangle & \frac{3}{5}\vert 10 \rangle + \frac{4}{5}\vert 11 \rangle \smallskip \\
			-\frac{12}{25}\vert 00 \rangle - \frac{16}{25}\vert 01 \rangle + \frac{9}{25}\vert 10 \rangle + \frac{12}{25}\vert 11 \rangle & \frac{16}{25}\vert 00 \rangle - \frac{12}{25}\vert 01 \rangle - \frac{12}{25}\vert 10 \rangle + \frac{9}{25}\vert 11 \rangle
		\end{pmatrix}$\smallskip \\
		$X''_{1,0}=
		\begin{pmatrix}
			-\frac{4}{5}\vert 00 \rangle + \frac{3}{5}\vert 01 \rangle & \frac{3}{5}\vert 00 \rangle + \frac{4}{5}\vert 01 \rangle \smallskip \\
			\frac{9}{25}\vert 00 \rangle + \frac{12}{25}\vert 01 \rangle + \frac{12}{25}\vert 10 \rangle + \frac{16}{25}\vert 11 \rangle & -\frac{12}{25}\vert 00 \rangle + \frac{9}{25}\vert 01 \rangle - \frac{16}{25}\vert 10 \rangle + \frac{12}{25}\vert 11 \rangle
		\end{pmatrix}$
	\end{center}
	Then the following  
	$$W_{a,b,d}=\begin{pmatrix} C_a &	X'_{0,0}&X'_{0,1}\\
		X''_{1,0}&C_b&X'_{1,1}\\
		X''_{0,0}&X'_{1,0}&C_d
	\end{pmatrix}$$ is a QLS$(6)$ with cardinality  $c=24+2|\{a,b,d\}|$, thereby $c\in \{26,28,30\}$.

	Clearly, $W_3=
	\begin{pmatrix}
		D_0&E_0\\
		E_0&F_1
	\end{pmatrix}$ and $W_4=
	\begin{pmatrix}
		D_0&E_0\\G_1&D_1
	\end{pmatrix}$ are QLS$(6)$ with cardinality $9,11$ respectively.  And 
	$W_5=
	\begin{pmatrix}
		A_0&B_0&C_0\\
		C_1&A_0&B_1\\
		B_2&C_2&A_1
	\end{pmatrix}$ is a QLS$(6)$ with cardinality $16$.  A QLS$(6)$ with cardinality $36$ exists by Theorem \ref{MQLS(n)}. \qed

\begin{lemma}\label{infiniteQSL(6)Cardinality36}
	For $i\in \mathbb{N}$, let  $\widetilde{W}_{i}$ be defined by Equation $(\ref{matrixtensorproduct})$ with $U_{i}$ and $V_0$, where  \begin{equation}\label{Vab}
			U_{i}=
			\begin{pmatrix}
				\frac{1}{\sqrt{{(3i+2)}^2 + 1}} ( \vert 0 \rangle + (3i+2) \vert 1 \rangle ) &	\frac{1}{\sqrt{(3i+2)^2 + 1}} ( (3i+2)\vert 0 \rangle - \vert 1 \rangle )\\
				\frac{1}{\sqrt{{(3i+3)}^2 + 1}} ( \vert 0 \rangle + (3i+3) \vert 1 \rangle ) &	\frac{1}{\sqrt{{(3i+3)}^2 + 1}} ( (3i+3)\vert 0 \rangle - \vert 1 \rangle ) \\
				\frac{1}{\sqrt{{(3i+4)}^2 + 1}} ( \vert 0 \rangle + (3i+4) \vert 1 \rangle ) &	\frac{1}{\sqrt{{(3i+4)}^2 + 1}} ( (3i+4)\vert 0 \rangle - \vert 1 \rangle )
			\end{pmatrix}.
	\end{equation}
	Then all  $\widetilde{W}_{i}$ are QLS$(6)$'s with maximal cardinality $36$ such that all $72$ elements of every two QLS$(6)$s are distinct.
\end{lemma}

\proof As described above, each $\widetilde{W}_{i}$ is a  QLS(6) with maximal cardinality $36$ by Lemma \ref{ProductConstruction}. For  $i\neq j$, since $i-j\neq 0$, $1+(3i+2)(3j+2)\neq 0$ and $1-(3i+2)(3j+2)\neq 0$, all $72$ elements of $\widetilde{W}_{i}$ and $\widetilde{W}_{j}$ are distinct. \qed

Define 
\begin{equation}\label{Vab}
	U_{0}=
	\begin{pmatrix}
		|0\rangle &|1\rangle\smallskip \\
		\frac{1}{5}(3|0\rangle + 4|1\rangle) 
		&\frac{1}{5}(4|0\rangle - 3|1\rangle) \smallskip\\
		\frac{1}{5}(4|0\rangle + 3|1\rangle) 
		&\frac{1}{5}(3|0\rangle - 4|1\rangle)
	\end{pmatrix},
\end{equation}
\begin{equation}\label{Vab}
	V_{0}=
	\begin{pmatrix}
		|0\rangle & |1\rangle & |2\rangle\smallskip \\
		\frac{1}{3}|0\rangle + \frac{2}{3}|1\rangle + \frac{2}{3}|2\rangle 
		&\frac{2}{3}|0\rangle + \frac{1}{3}|1\rangle - \frac{2}{3}|2\rangle 
		&\frac{2}{3}|0\rangle - \frac{2}{3}|1\rangle + \frac{1}{3}|2\rangle
	\end{pmatrix}.
\end{equation}
Then $U_{0}$ is a row-quantum Latin rectangle over $\mathcal{H}_2$ with cardinality $6$. By applying Lamma \ref{ProductConstruction} with $U_{0}$ and $V_{0}$, the $6\times 6$ matrix $W_0$ defined by Equation $(\ref{matrixtensorproduct})$ is a QLS$(6)$ with maximal cardinality 36, which is displayed below for use later,
\begin{equation}\label{Wab}
	W_{0}=\begin{pmatrix}
		(|00\rangle, |01\rangle, |02\rangle,|10\rangle, |11\rangle,|12\rangle)X_{1}\\
		(|00\rangle, |01\rangle, |02\rangle,|10\rangle, |11\rangle,|12\rangle)X_{2}\\
		(|00\rangle, |01\rangle, |02\rangle,|10\rangle, |11\rangle,|12\rangle)X_{3}\\
		(|00\rangle, |01\rangle, |02\rangle,|10\rangle, |11\rangle,|12\rangle)X_{4}\\	
		(|00\rangle, |01\rangle, |02\rangle,|10\rangle, |11\rangle,|12\rangle)X_{5}\\	
		(|00\rangle, |01\rangle, |02\rangle,|10\rangle, |11\rangle,|12\rangle)X_{6}\\	
	\end{pmatrix},	
\end{equation}
where $X_{i}$ are as follows:
\[
X_1 = \begin{pmatrix}
	0 & 0 & 0 & \frac{2}{3} & \frac{2}{3} & \frac{1}{3} \smallskip \\
	0 & 0 & 0 & \frac{1}{3} & -\frac{2}{3} & \frac{2}{3} \smallskip \\
	0 & 0 & 0 & -\frac{2}{3} & \frac{1}{3} & \frac{2}{3} \\
	0 & 0 & 1 & 0 & 0 & 0 \\
	1 & 0 & 0 & 0 & 0 & 0 \\
	0 & 1 & 0 & 0 & 0 & 0 \\
\end{pmatrix},
\]
\[
X_2 = \begin{pmatrix}
	0 & 0 & 1 & 0 & 0 & 0 \\
	1 & 0 & 0 & 0 & 0 & 0 \\
	0 & 1 & 0 & 0 & 0 & 0 \\
	0 & 0 & 0 & \frac{2}{3} & \frac{2}{3} & \frac{1}{3} \smallskip \\
	0 & 0 & 0 & \frac{1}{3} & -\frac{2}{3} & \frac{2}{3} \smallskip\\
	0 & 0 & 0 & -\frac{2}{3} & \frac{1}{3} & \frac{2}{3} \\
\end{pmatrix},
\]
\[
X_3 = \begin{pmatrix}
	0 & \frac{4}{5} & 0 & \frac{2}{5} & \frac{1}{5} & \frac{2}{5} \smallskip \\
	0 & 0 & \frac{4}{5} & -\frac{2}{5} & \frac{2}{5} & \frac{1}{5} \smallskip \\
	\frac{4}{5} & 0 & 0 & \frac{1}{5} & \frac{2}{5} & -\frac{2}{5} \smallskip \\
	0 & -\frac{3}{5} & 0 & \frac{8}{15} & \frac{4}{15} & \frac{8}{15} \smallskip \\
	0 & 0 & -\frac{3}{5} & -\frac{8}{15} & \frac{8}{15} & \frac{4}{15} \smallskip \\
	-\frac{3}{5} & 0 & 0 & \frac{4}{15} & \frac{8}{15} & -\frac{8}{15} \\
\end{pmatrix},
\]
\[
X_4 = \begin{pmatrix}
	0 & \frac{3}{5} & 0 & \frac{8}{15} & \frac{4}{15} & \frac{8}{15} \smallskip\\
	0 & 0 & \frac{3}{5} & -\frac{8}{15} & \frac{8}{15} & \frac{4}{15} \smallskip\\
	\frac{3}{5} & 0 & 0 & \frac{4}{15} & \frac{8}{15} & -\frac{8}{15} \smallskip\\
	0 & \frac{4}{5} & 0 & -\frac{2}{5} & -\frac{1}{5} & -\frac{2}{5} \smallskip\\
	0 & 0 & \frac{4}{5} & \frac{2}{5} & -\frac{2}{5} & -\frac{1}{5} \smallskip\\
	\frac{4}{5} & 0 & 0 & -\frac{1}{5} & -\frac{2}{5} & \frac{2}{5} \\
\end{pmatrix},
\]
\[
X_5 = \begin{pmatrix}
	\frac{3}{5} & 0 & 0 & \frac{4}{15} & \frac{8}{15} & \frac{8}{15} \smallskip\\
	0 & \frac{3}{5} & 0 & \frac{8}{15} & \frac{4}{15} & -\frac{8}{15} \smallskip\\
	0 & 0 & \frac{3}{5} & \frac{8}{15} & -\frac{8}{15} & \frac{4}{15} \smallskip\\
	-\frac{4}{5} & 0 & 0 & \frac{1}{5} & \frac{2}{5} & \frac{2}{5} \smallskip\\
	0 & -\frac{4}{5} & 0 & \frac{2}{5} & \frac{1}{5} & -\frac{2}{5} \smallskip\\
	0 & 0 & -\frac{4}{5} & \frac{2}{5} & -\frac{2}{5} & \frac{1}{5} \\
\end{pmatrix},
\]
\[
X_6 = \begin{pmatrix}
	\frac{4}{5} & 0 & 0 & \frac{1}{5} & \frac{2}{5} & \frac{2}{5} \smallskip\\
	0 & \frac{4}{5} & 0 & \frac{2}{5} & \frac{1}{5} & -\frac{2}{5} \smallskip\\
	0 & 0 & \frac{4}{5} & \frac{2}{5} & -\frac{2}{5} & \frac{1}{5} \smallskip\\
	\frac{3}{5} & 0 & 0 & -\frac{4}{15} & -\frac{8}{15} & -\frac{8}{15} \smallskip\\
	0 & \frac{3}{5} & 0 & -\frac{8}{15} & -\frac{4}{15} & \frac{8}{15} \smallskip\\
	0 & 0 & \frac{3}{5} & -\frac{8}{15} & \frac{8}{15} & -\frac{4}{15} \\
\end{pmatrix}.
\]
Note that each $X_{i}$ is an orthonormal matrix for $1\leq i\leq 6$ and all of the $j$th column vectors of $X_{i}$, $1\leq i\leq 6$, form an orthonormal basis in $\mathbb{R}^6$ for $1\leq j\leq 6$.

Define 
\begin{equation}\label{J1}
J_1 = \begin{pmatrix} 
	1 & 0 & 0 & 0 & 0 & 0 \\
	0 & \frac{2}{3} & -\frac{2}{3} & \frac{1}{3} & 0 & 0 \smallskip \\
	0 & \frac{1}{3} & \frac{2}{3} & \frac{2}{3} & 0 & 0 \smallskip \\
	0 & \frac{2}{3} & \frac{1}{3} & -\frac{2}{3} & 0 & 0 \\
	0 & 0 & 0 & 0 & \frac{3}{5} & -\frac{4}{5} \smallskip \\
	0 & 0 & 0 & 0 & \frac{4}{5} & \frac{3}{5} \\
\end{pmatrix},
\end{equation}
\begin{equation}\label{J2}
	J_2 = \begin{pmatrix}
	1 & 0 & 0 & 0 & 0 & 0 \\
	0 & 1 & 0 & 0 & 0 & 0 \\
	0 & 0 & \frac{1}{2} & \frac{1}{2} & \frac{1}{2} & \frac{1}{2} \smallskip \\
	0 & 0 & \frac{1}{2} & -\frac{1}{2} & \frac{1}{2} & -\frac{1}{2} \smallskip \\
	0 & 0 & \frac{1}{2} & \frac{1}{2} & -\frac{1}{2} & -\frac{1}{2} \smallskip \\
	0 & 0 & \frac{1}{2} & -\frac{1}{2} & -\frac{1}{2} & \frac{1}{2} \\
	\end{pmatrix},
\end{equation}

\begin{equation}\label{J3}
	J_3=\begin{pmatrix}
		1 & 0 & 0 & 0 & 0 & 0 \\
		0 & 1 & 0 & 0 & 0 & 0 \\
		0 & 0 & 1 & 0 & 0 & 0 \\
		0 & 0 & 0 & \frac{2}{3} & -\frac{2}{3} & \frac{1}{6} \smallskip \\
		0 & 0 & 0 & \frac{1}{3} & \frac{2}{3} & \frac{2}{3} \smallskip \\
		0 & 0 & 0 & \frac{2}{3} & \frac{1}{3} & -\frac{2}{5} \\
	\end{pmatrix},
\end{equation}

\begin{equation}\label{J4}
	J_4=\begin{pmatrix}
	1 & 0 & 0 & 0 & 0 & 0 \\
	0 & \frac{2}{3} & -\frac{2}{3} & \frac{1}{3} & 0 & 0 \smallskip \\
	0 & \frac{1}{3} & \frac{2}{3} & \frac{2}{3} & 0 & 0 \smallskip \\
	0 & \frac{2}{3} & \frac{1}{3} & -\frac{2}{3} & 0 & 0  \\
	0 & 0 & 0 & 0 & -\frac{4}{5} & \frac{3}{5} \smallskip \\
	0 & 0 & 0 & 0 & \frac{3}{5} & \frac{4}{5} \\
	\end{pmatrix}.
\end{equation}

For $k\in \{1,2,3,4\}$, define
\begin{equation}\label{W0}
	M_{k}=\begin{pmatrix}
		(|00\rangle, |01\rangle, |02\rangle,|10\rangle, |11\rangle,|12\rangle)X_1J_kX_1^{-1}X_1\\
		(|00\rangle, |01\rangle, |02\rangle,|10\rangle, |11\rangle,|12\rangle)X_1J_kX_1^{-1}X_2\\
		(|00\rangle, |01\rangle, |02\rangle,|10\rangle, |11\rangle,|12\rangle)X_1J_kX_1^{-1}X_3\\
	    (|00\rangle, |01\rangle, |02\rangle,|10\rangle, |11\rangle,|12\rangle)X_1J_kX_1^{-1}X_4\\	
	    (|00\rangle, |01\rangle, |02\rangle,|10\rangle, |11\rangle,|12\rangle)X_1J_kX_1^{-1}X_5\\	
	    (|00\rangle, |01\rangle, |02\rangle,|10\rangle, |11\rangle,|12\rangle)X_1J_kX_1^{-1}X_6\\	
	\end{pmatrix}.	
\end{equation}
Since all $X_1, J_k$ are orthgonormal matrices, the vectors $(|00\rangle, |01\rangle, |02\rangle, |10\rangle,|11\rangle,|12\rangle)X_1J_kX_1^{-1}$ also form an orthonormal basis for $\mathcal{H}_2\otimes \mathcal{H}_3$. Since the column vectors of $X_1, X_2, X_3, X_4, X_5, X_6$ form a QLS$(6)$ with cardinality 36 in $\mathbb{C}^6$, each $M_k$ is a QLS$(6)$ with cardinality 36. Further, simple computation shows the following facts:

(P1) $W_0$ and $M_1$ have a common element, 

(P2) $W_0$ and $M_2$ have exactly two common elements, 

(P3) $W_0$ and $M_3$ have exactly six common elements,

(P4) $M_1$ and $M_4$ have exactly thirteen common elements.

\noindent Hence, there is a QLS$(6)$  with $\ell$ distinct elements not from $W_0$ for $\ell\in \{30,34,35\}$. Since the coefficients of $\widetilde{W}_{i}$ defined in Lemma \ref{infiniteQSL(6)Cardinality36} are irrational in orthonormal basis $|00\rangle, |01\rangle, |02\rangle, |10\rangle, |11\rangle, |12\rangle$, there are 36 distinct elements in $\widetilde{W}_{i}$ not from $W_0$.

\begin{lemma}\label{H0H1}
Define $H_0$,$H_1$ with cardinality $6,14$ as follows:\\
\begin{center}
$H_0=\begin{pmatrix}
A_0&B_0&C_0\\
B_0&C_0&A_0\\
C_0&A_0&B_0\\
\end{pmatrix}$,
$H_1=\begin{pmatrix}
	A_0 & B_0& C_0\\C_0&A_1&B_0\\
	B_1 &C_1 & A_2\\
\end{pmatrix}$,\\
\end{center}
where $A_a,B_b$ and $C_c$ are defined in (\ref{A_aB_a}). 
There is a QLS$(6)$  $H_{\ell}$ with  $\ell$ distinct elements neither from $H_0$ nor from $H_1$ for $\ell\in  \{2,3,4,5,6,7,8,9,10,11,12,13,14,16,18,20,22,24,26,36\} $.
There is a QLS$(6)$ $H_{\ell}'$ with $\ell$ distinct elements not from $W_0$ in $(\ref{Wab})$ for 
$\ell\in \{2,3,4,5,6,7,8,9,10,11,12,13,14,16,18,20,22,36\}$. Furthermore, there are infinite number of QLS$(6)$s $\widetilde{W}_{i}$ in Lemma $\ref{infiniteQSL(6)Cardinality36}$, each with 36 distinct elements neither from $H_{\ell}$ nor from $H_{\ell}'$. 
\end{lemma}

\proof In the proof of Lemma \ref{lemma:QLS(6)}, we can  choose an appropriate  QLS(4) $X$ and some $a,b,d$ to get a QLS$(6)$ $L_{a,b,d}^i$.
If we choose $X=
\begin{pmatrix}
	A_0&B_0\\
	B_0&A_0\\
\end{pmatrix}$, then
$L_{2,0,0}^0,L_{2,3,0}^0,L_{2,3,4}^0$ have 
$2,4,6$ distinct elements not from $H_0$ or $H_1$, respectively. 
If $X=\begin{pmatrix}
	A_3&B_2\\
	B_3&A_4\\
\end{pmatrix}$, then $L_{0,0,0}^1,L_{2,0,0}^1,L_{2,3,0}^1,L_{2,3,4}^1$ have exactly $8,10,12,14$ distinct elements not from $H_0$,  $H_1$, or $W_1$ respectively.
If $X=X''$, then $L_{0,0,0}^2,L_{2,0,0}^2,L_{2,3,0}^2,L_{2,3,4}^2$ have $16,18,20,22$ distinct elements not from $H_0$, $H_1$,
or $W_1$ respectively. Also, $W_{2,3,0},W_{2,3,4}$ in the proof of Lemma \ref{lemma:QLS(6)} have $24,26$ distinct elements not from $H_0$ or $H_1$.

\begin{equation}\label{S1}
	S=\begin{pmatrix}
		1 & 0 & 0 & 0 \\
		0 & \frac{1}{\sqrt{3}} & -\frac{2}{\sqrt{3}} & \frac{2}{\sqrt{3}} \smallskip \\
		0 & \frac{2}{\sqrt{3}} & -\frac{1}{\sqrt{3}}\omega & -\frac{2}{\sqrt{3}}\omega^2 \smallskip \\
		0 & \frac{2}{\sqrt{3}} & \frac{2}{\sqrt{3}}\omega^2 & \frac{1}{\sqrt{3}} \omega
	\end{pmatrix}.
\end{equation}
where $\omega$ is a primitive cubic root of unity.

 Clearly, $S$ is an orthonormal matrix. It follows that the vectors $\alpha_1,\alpha_2,\alpha_3,\alpha_4$ defined by $(\alpha_1,\alpha_2,\alpha_3,\alpha_4)=(|00\rangle, |01\rangle, |10\rangle,|11\rangle)S$ also form an orthonormal basis for $\mathcal{H}_2^{\otimes 2}$. Note that $\alpha_1=|00\rangle$.
If $X=\begin{pmatrix}
	|00\rangle& \alpha_2& \alpha_3& \alpha_4\\
	\alpha_2& \alpha_3& \alpha_4 &|00\rangle  \smallskip \\
\alpha_3& \alpha_4 &|00\rangle &\alpha_2  \smallskip \\
	\alpha_4 & |00\rangle& \alpha_2 & \alpha_3
\end{pmatrix}$, then $L_{0,0,0}^3,L_{2,0,0}^3,L_{2,3,0}^3,L_{2,3,4}^3$ defined in the proof of Lemma \ref{QLS(6)} have $3,5,7,9$ distinct elements not from $H_0$,  $H_1$ or $W_1$ respectively.
If $X=\begin{pmatrix}
	|00\rangle& \alpha_2& \alpha_3& \alpha_4\\
	\alpha_2& |00\rangle& \alpha_4 & \alpha_3 \smallskip \\
	\frac{\alpha_3+\alpha_4}{\sqrt{2}}&\frac{\alpha_3-\alpha_4}{\sqrt{2}} &\frac{|00\rangle+\alpha_2}{\sqrt{2}} &\frac{|00\rangle-\alpha_2}{\sqrt{2}} \smallskip \\
		\frac{\alpha_3+\alpha_4}{\sqrt{2}} & 	\frac{\alpha_3-\alpha_4}{\sqrt{2}}& \frac{|00\rangle-\alpha_2}{\sqrt{2}} & \frac{|00\rangle+\alpha_2}{\sqrt{2}}
\end{pmatrix}$, then $L_{2,3,0}^4,L_{2,3,4}^4$ defined in the proof of Lemma \ref{QLS(6)} have $11,13$ distinct elements not from $H_0$, $H_1$ or $W_1$, respectively.

Clearly, there exists a QLS$(6)$ with cardinality 36 and these 36 distinct states differ from $H_0$, $H_1$ or $W_1$ by Lamma \ref{infiniteQSL(6)Cardinality36}.   

By comparing the coordinate vectors of elements in $\widetilde{W}_{i}$  and $L_{a,b,d}^j$, it is easy to check that there are $m-1$ QSL$(6)$, say $\widetilde{W}_{i}$ ($5\leq i\leq m+4$) with cardinality 36 and these $36m-36$ distinct elements differ from those in $L_{a,b,d}^j$, $H_0$ and $H_1$. \qed

 We are in a position to show our main result of this note.

\begin{lemma}
	For $m\geq 2$ and any $c\in [6m,36m^2]\setminus \{6m+1\}$, there is a QLS$(6m)$  with cardinality $c$.
\end{lemma}
\proof For $m=2$, the conclusion holds by Theorem \ref{QLS(4m)}. For $m\geq 3$, let $A=(a_{i,j})$, where $a_{i,j}=j-i\pmod m$. Then $A$ is a classical Latin square over $\{0,1,\ldots,m-1\}$. 
Repalcing each $a_{i,j}$ with $|a_{i,j}\rangle \otimes Y_{i,j}$ where $Y_{i,j}$ is a QLS$(6)$, we obtain a QLS$(6m)$ in $\mathcal{H}_m\otimes \mathcal{H}_2\otimes \mathcal{H}_3$.
\begin{center}
	$\begin{pmatrix}
		|0\rangle\otimes Y_{0,0} & |1\rangle\otimes Y_{0,1}&|2\rangle\otimes Y_{0,2}& \cdots &|m-1\rangle\otimes Y_{0,m-1} \\
		|m-1\rangle\otimes Y_{1,0} & |0\rangle\otimes Y_{1,1}&|1\rangle\otimes Y_{1,2}& \cdots &|m-2\rangle\otimes Y_{1,m-1}\\
		|m-2\rangle\otimes Y_{2,0} & |m-1\rangle\otimes Y_{2,1}&|0\rangle\otimes Y_{2,2}& \cdots &|m-3\rangle\otimes Y_{2,m-1}\\
		\vdots & \vdots& \vdots& \ddots & \vdots
		\\|1\rangle\otimes Y_{m-1,0}&|2\rangle\otimes Y_{m-1,1}&|3\rangle\otimes Y_{m-1,2}& \cdots &|0\rangle\otimes Y_{m-1,m-1}
	\end{pmatrix}$
\end{center}
In order that this QLS$(6m)$ has cardinality $c$, we have to choose appropriate QLS$(6)$'s $Y_{i,j}$. 

By Lemma \ref{H0H1}, there is a QLS$(6)$ $H_{\ell}$ with $\ell$ distinct elements neither from $H_0$ nor from $H_1$ for $\ell \in \{2,3,4,5,6,7,8,9,10,11,12,13,14,16,18,20,22,24,26,36\}$. Also, there are $m-1$ QLS$(6)$ $\widetilde{W}_{i}$ with cardinality 36 and these $36m-36$ distinct elements differ from those in $H_0, H_1$ and $H_{\ell}$.  

For $0\leq j\leq m-1$, take
$$Y_{i,i+j\pmod m}\in \left \{ \begin{array}{ll}
\{H_0,H_1\} & \text{if}\ i=0\\
 \{H_0,\widetilde{W}_{i},H_{\ell}\} & \text{if}\ i=1\\
\{Y_{0,j},\widetilde{W}_{i}\} & \text{if}\ i\geq 2.\\
\end{array} \right . $$
Then all $|a_{i,i+j}\rangle \otimes Y_{i,i+j}$, $0\leq i\leq m-1$, contribute to $6+8y_{j,0}+y_{j,1}+36y_{j,2}+\ldots+36y_{j,m-1}$ distinct elements where $y_{j,0}\in \{0,1\}$, $y_{j,1}\in \{0,2,3,4,5,6,7,8,9,10,11,12,13,14,16,18,20,22,24,26,36\}$ and $y_{j,i}\in \{0,1\}$ for $2\leq i\leq m-1$. Hence, this QLS$(6m)$ has totally
$$6m+8\sum_{j=0}^{m-1}y_{j,0}+\sum_{j=0}^{m-1}y_{j,1}+36\sum_{j=0}^{m-1}\sum_{i=2}^{m-1}y_{j,i}$$
distinct elements. Note that $\sum_{j=0}^{m-1}y_{j,0}$ runs through $[0,m]$, $\sum_{j=0}^{m-1}y_{j,1}$ runs through $[0,36m-18]\setminus \{1,36m-19,36m-21\}$ and $\sum_{j=0}^{m-1}\sum_{i=2}^{m-1}y_{j,i}$ runs through $[0,m(m-2)]$. Hence, for $c\in [6m,36m^2-22m-18]\setminus \{6m+1,36m^2-22m-19,36m^2-22m-21\}$, there is a QLS$(6m)$ with cardinality $c$.

By Lemma \ref{H0H1}, there is a QLS$(6)$ $H_{\ell}'$ with $\ell$ distinct elements not from $W_0$ for $\ell \in \{2,3,4,5,6,7,8,9$, $10,11,12,13,14,16,18,20,22,36\}$. Also, there are $m-1$ QLS$(6)$ $\widetilde{W}_{i}$ with cardinality 36 and these $36m-36$ distinct elements differ from those in $W_0$ and $H_{\ell}'$.  

For $0\leq j\leq m-1$, take
$$Y_{i,i+j\pmod m}\in \left \{ \begin{array}{ll}
	\{W_0\} & \text{if}\ i=0\\
	\{H_{\ell}',W_0,M_{1},M_2,M_{3},M_{4},\widetilde{W}_{i}\} & \text{if}\ i=1\\
	\{\widetilde{W}_{i}\} & \text{if}\ i\geq 2.\\
\end{array} \right . $$
Then all $|a_{i,i+j}\rangle \otimes Y_{i,i+j}$, $0\leq i\leq m-1$, contribute to $36+y_{j,1}+36(m-2)$ distinct elements where $y_{j,1}\in
 \{0,2,3,4,5,6,7,8,9,10,11,12,13,14,16,18,20,22,30,34,35,36\}$. Hence, this QLS$(6m)$ has totally
$$36m(m-1)+\sum_{j=0}^{m-1}y _{j,1}$$
distinct elements.

For $m=2$, there is a QLS$(12)$ with cardinaltiy $c$ for any $c\in [12,144]\setminus \{13\}$ \cite{ZCPreprint}.

For $m=3$,  note that $\sum_{j=0}^{m-1}y_{j,1}$ runs through $[0,36m]\setminus\{1,36m-11\}$. Hence, for $c\in [36m^2-36m,36m^2]\setminus\{36m^2-36m+1,36m^2-11\}$, there is a QLS$(6m)$ with cardinality $c$. 
Clearly, $36m^2-36m<36m^2-22m-21$,  thereby, there is a QLS$(6m)$ with cardinaltiy $c$ for any $c\in [6m,36m^2]\setminus \{6m+1,36m^2-11\}$.

The following QLS$(18)$ has cardinality 313.

\begin{center}
$\begin{pmatrix}
		|0\rangle \otimes W_0 & |1\rangle \otimes W_0 & |2\rangle \otimes W_0\\
		|2\rangle \otimes M_3 & |0\rangle \otimes M_2 & |1\rangle \otimes M_2\\
		|1\rangle \otimes \widetilde{W}_{5} & |2\rangle \otimes \widetilde{W}_{5} & |0\rangle \otimes M_{1}
	\end{pmatrix}
	$
\end{center}

For $m\geq 4$, note that $\sum_{j=0}^{m-1}x_{j,1}$ runs through $[0,36m]\setminus \{1\}$. Hence, for $c\in [36m^2-36m,36m^2]\setminus \{36m^2-36m+1\}$, there is a QLS$(6m)$ with cardinality $c$.
Clearly, $36m^2-36m<36m^2-22m-21$, thereby there is a QLS$(6m)$ with cardinaltiy $c$ for any $c\in [6m,36m^2]\setminus \{6m+1\}$.

This completes the proof. \qed

\end{document}